\documentclass[aps,pra,twocolumn,showpacs,superscriptaddress,floatfix]{revtex4-1}

\usepackage{amsfonts,amsmath,amssymb}
\usepackage{graphicx}

\newcommand{\ket}[1]{|#1\rangle}
\newcommand{\moy}[1]{\langle#1\rangle}
\newcommand{\Exp}[1]{\mathrm{e}^{#1}}

\begin{document}

\title{Perfect squeezing by damping modulation in circuit quantum electrodynamics}

\author{Nicolas Didier}
\affiliation{D\'epartment de Physique, Universit\'e de Sherbrooke, Sherbrooke, Qu\'ebec J1K 2R1, Canada}
\affiliation{Department of Physics, McGill University, Montreal, Quebec H3A 2T8, Canada}
\author{Farzad Qassemi}
\affiliation{D\'epartment de Physique, Universit\'e de Sherbrooke, Sherbrooke, Qu\'ebec J1K 2R1, Canada}
\author{Alexandre Blais}
\affiliation{D\'epartment de Physique, Universit\'e de Sherbrooke, Sherbrooke, Qu\'ebec J1K 2R1, Canada}

\begin{abstract}
Dissipation-driven quantum state engineering uses the environment to steer the state of quantum systems and preserve quantum coherence in the steady state. We show that modulating the damping rate of a microwave resonator generates a vacuum squeezed state of arbitrary squeezing strength, thereby constituting a mechanism allowing perfect squeezing. Given the recent experimental realizations in circuit QED of a microwave resonator with a tunable damping rate [Yin {\it et al.}, Phys. Rev. Lett.~{\bf 110}, 107001 (2013)], superconducting circuits are an ideal playground to implement this technique. By dispersively coupling a qubit to the microwave resonator, it is possible to obtain qubit-state dependent squeezing.
\end{abstract}

\pacs{
42.50.Dv, 
03.65.Yz, 
03.67.Lx. 
}

\maketitle

\textit{Introduction} -- 
The environment surrounding a quantum system induces decoherence and destroys quantum correlations. Quantum state engineering and manipulation is then limited by the coherence lifetime of the system. Surprisingly, if the coupling to the environment is controlled, it can be used as a resource to steer the system to a desired, fully coherent, quantum state. For example, it has been theoretically shown that this type of quantum bath engineering can be used for universal quantum computation~\cite{verstraete:2009a,kastoryano:2013a} and to create robust quantum memories~\cite{pastawski:2011a}. Possible realizations of quantum bath engineering have also been theoretically explored in the context of optical cavities~\cite{kastoryano:2011a,sweke:2013a}, optomechanical systems~\cite{wang:2013a}, and circuit quantum electrodynamics (QED)~\cite{reiter:2013a,leghtas:2013a,reiter:2013a}. Experimentally, dissipation-driven steady state entanglement has been obtained with atomic clouds~\cite{krauter:2011a}, trapped ions~\cite{lin:2013a}, and superconducting circuits~\cite{shankar:2013a}. The preparation via damping of arbitrary superpositions in the state of a superconducting qubit has also been realized~\cite{murch:2012a}. 

Because of the improved measurement sensitivity that they provide, squeezed states are particularly interesting quantum states to stabilize~\cite{giovannetti:2004a}. In the optical domain, the standard approach to prepare such states is to pump a cavity containing a Kerr medium at twice the cavity frequency~\cite{gardiner:2004b}. With this coherent approach, the variance of the intracavity squeezed quadrature  is however at most reduced by a factor of 2. This is the well known 3$\,$dB limit for squeezing
\footnote{
Under strong driving at a frequency $\omega_p$, a cavity of frequency $\omega_r$ containing a Kerr medium can be described by $H=\omega_r \hat{a}^\dag \hat{a} + iK(\hat{a}^{\dag2} \mathrm{e}^{-i \omega_p t} - \hat{a}^2\mathrm{e}^{i \omega_p t})$. In the steady state and with $\omega_p = 2\omega_r$, this Hamiltonian leads to squeezed quadratures of variance $\Delta X^2=1/(1+\mu)$ and $\Delta Y^2=1/(1-\mu)$, where $\mu=4K/\kappa <1$. Squeezing is however not ideal since $\Delta X^2\Delta Y^2=1/(1-\mu^2)>1$, and is limited by $\Delta X^2\geq1/2$.
}. 
Moreover the product of the variance of the two quadratures is larger than the Heisenberg limit; that is, the squeezing is not ideal.

Motivated by a recent experiment in the field of circuit QED~\cite{yin:2013a}, we study the effect of periodically modulating the coupling between a microwave resonator and its environment, and we show that dissipation-driven squeezing can go beyond the limits encountered with the coherent generation of squeezed states. Indeed, we show that when the coupling strength of a microwave resonator to its environment is modulated at twice the resonator frequency $\omega_r$, the intra-resonator field can be ideally squeezed. The squeezed quadrature is arbitrarily reduced down to zero, beating the standard 3$\,$dB limit and allowing perfect squeezing. The degree and axis of squeezing are controlled by the modulation amplitude and phase. With a qubit dispersively coupled to the resonator, we show that it is possible to obtain qubit-state dependent squeezing and discuss implications for qubit readout. Preparation of squeezed states by dissipation has also been explored theoretically with atomic ensembles to generate spin squeezing~\cite{dalla-torre:2013a}, in optomechanics to obtain two-mode mechanical squeezing~\cite{tan:2013a}, using back-action evading measurement with stroboscopic observations~\cite{braginsky:1980,*caves:1980,*clerk:2008} and with multichromatic excitations of trapped ions~\cite{cirac:1993} or nano-mechanical resonators~\cite{rabl:2004}.

In cavity QED, modulating the damping rate would correspond to changing the transparency of the mirrors forming the cavity. The equivalent control with microwave resonators has already been reported in Ref.~\cite{yin:2013a}. In that experiment, the damping rate $\kappa$ of a $\lambda/4$ resonator is controlled using an external magnetic flux coupled to a variable inductance which is itself based on a SQUID loop~\cite{bialczak:2011a}. Using this approach, the damping rate was abruptly changed from zero to a rate 1000 times the intrinsic resonator $\kappa$ in a few nanoseconds. Here, we propose to modulate damping at a frequency $2\omega_r$ for which there is enough bandwidth. The fact that it should be possible to produce squeezed microwave light by periodically pumping a SQUID coupled to a resonator is certainly not surprising and has already been experimentally reported numerous times~\cite{yamamoto:2008a,castellanos-beltran:2008a,vijay:2011a}. As we argue below, the present approach however represents a distinct squeezing mechanism. Before moving to a more complete description, it is useful to have an intuitive picture for this phenomenon in the classical regime. Indeed, by modulating the damping at twice the mode frequency, the amplitude of one of the quadratures is in phase with the modulation while the other is out of phase. As a result, only the former easily leaks out of the resonator and  the light field is squashed.

\textit{Squeezing by dissipation} -- 
To describe squeezing by dissipation in the quantum regime, we obtain a master equation for the damping modulated resonator essentially following the standard Born-Markov approach~\cite{gardiner:2004b}. The environment is modeled by an infinite number of harmonic oscillators, which we take to be at zero temperature for simplicity. The resonator of frequency $\omega_r$ is described by the creation operator $\hat{a}^\dag$, while $\hat{f}^\dag(\omega)$ creates an excitation of frequency $\omega$ in the environment. The interaction between the resonator and the environment is characterized by the coupling constant $u(\omega)$ and the bosonic density $d(\omega)$. This interaction is time-modulated with the general Fourier expansion
\begin{equation}
\lambda(t)=\sum_n\lambda_n\,\Exp{i\omega_nt}.
\end{equation}
As discussed in more detail in the Appendix~\ref{AppA}, such a dynamical coupling to the environment is described by the Hamiltonian
\begin{equation}
H_\mathrm{int}=\{[1+\lambda^*(t)]\hat{F}^\dag+[1+\lambda(t)]\hat{F}\}(\hat{a}^\dag+\hat{a}),
\end{equation}
where $\hat{F}^\dag \equiv \int_0^\infty\mathrm{d}\omega\sqrt{d(\omega)}u^*(\omega)\hat{f}^\dag(\omega)$.
Tracing over the environmental degrees of freedom using the standard approximations (see Appendix~\ref{AppA}), we find that the system is described by the master equation (ME) $\dot{\rho}=-i[\omega_r\hat{a}^\dag\hat{a},\rho]+\mathcal{L}\rho$, where the Lindbladian $\mathcal{L}\cdot$ is composed of ordinary dissipators $\mathcal{D}[\hat{a}]\cdot=\hat{a}\cdot\hat{a}^\dag-\frac{1}{2}\{\hat{a}^\dag\hat{a},\cdot\}$ and, interestingly, of squeezing superoperators
\begin{equation}
\mathcal{S}[\hat{a}]\cdot=\hat{a}\cdot\hat{a}-\tfrac{1}{2}\{\hat{a}^2,\cdot\}.
\label{squeezingsuperoperator}
\end{equation}

Without external modulation, the squeezing superoperator averages out and can be eliminated using the rotating wave approximation while deriving the ME. The modulation of the coupling at twice the resonator frequency activates the effect of $\mathcal{S}\cdot$.
This can be easily understood from Eq.~\eqref{squeezingsuperoperator} since, in the interaction picture, $\hat{a}^2$ rotates at $2\omega_r$. More precisely, $\mathcal{S}\cdot$ is relevant in the ME if in the interaction picture its terms oscillate slower than the prefactor of $\mathcal{S}\cdot$ itself. In practice, we thus require that $|\omega_n-2\omega_r|\lesssim\kappa$, with $\kappa$ the resonator damping without modulations. In this situation, the Lindbladian reads (see Appendix~\ref{AppA})
\begin{equation}
\mathcal{L}=\kappa \mathcal{D}[\hat{a}]+|\lambda(t)|^2\kappa \mathcal{D}[\hat{a}^\dag]+\lambda(t)\kappa \mathcal{S}[\hat{a}]+\lambda^*(t)\kappa \mathcal{S}[\hat{a}^\dag].
\label{genL}
\end{equation}

It is worth noting that 
this squeezing mechanism is not simply related to the standard degenerate parametric amplifier Hamiltonian used to generate squeezing based on a coherent process. Indeed, defining $\rho' = S^\dag \rho S$, with $S=\exp\{(\xi^*\hat{a}^2-\xi\hat{a}^{\dag2})/2\}$ the squeezing operator, yields a ME corresponding to a linear oscillator damped at a renormalized rate $\Gamma$ by a zero-temperature bath: $\dot\rho' = -i[\omega_r\hat{a}^\dag\hat{a},\rho'] + \Gamma\mathcal{D}[\hat{a}]\rho'$ (see Appendix~\ref{AppB}). Back in the laboratory frame, the ground state of this equation corresponds to a vacuum squeezed state, illustrating clearly the role of damping. We also note that modulating the resonator frequency at twice its bare frequency $\omega_r$ would also activate terms of the form $\mathcal{S}\cdot$ in the ME (see Appendix~\ref{AppD}). For the realization of Ref.~\cite{yin:2013a}, such a modulation is present but with a very small relative amplitude $\delta\omega_r/\omega_r \sim 0.2~\%$, and this contribution can be safely neglected. Terms of the form $\mathcal{S}\cdot$ would also be present if the cavity were subjected to broadband squeezed noise~\cite{gardiner:2004b}.  Finally, and as discussed in more detail in the Appendix~\ref{AppA}, we point out that the Lindbladian of Eq.~\eqref{genL} is obtained after eliminating the density of the environmental modes at the frequency~$3\omega_r$. This can be efficiently achieved using a Purcell filter~\cite{reed:2010a}. As discussed below, without this additional filtering, the squeezing that can be obtained is degraded (but not eliminated). Nonradiative damping will also lead to non-ideal squeezing. In practice this will be a small correction for overcoupled resonators.  

\begin{figure}
\centering
\includegraphics[width=\columnwidth]{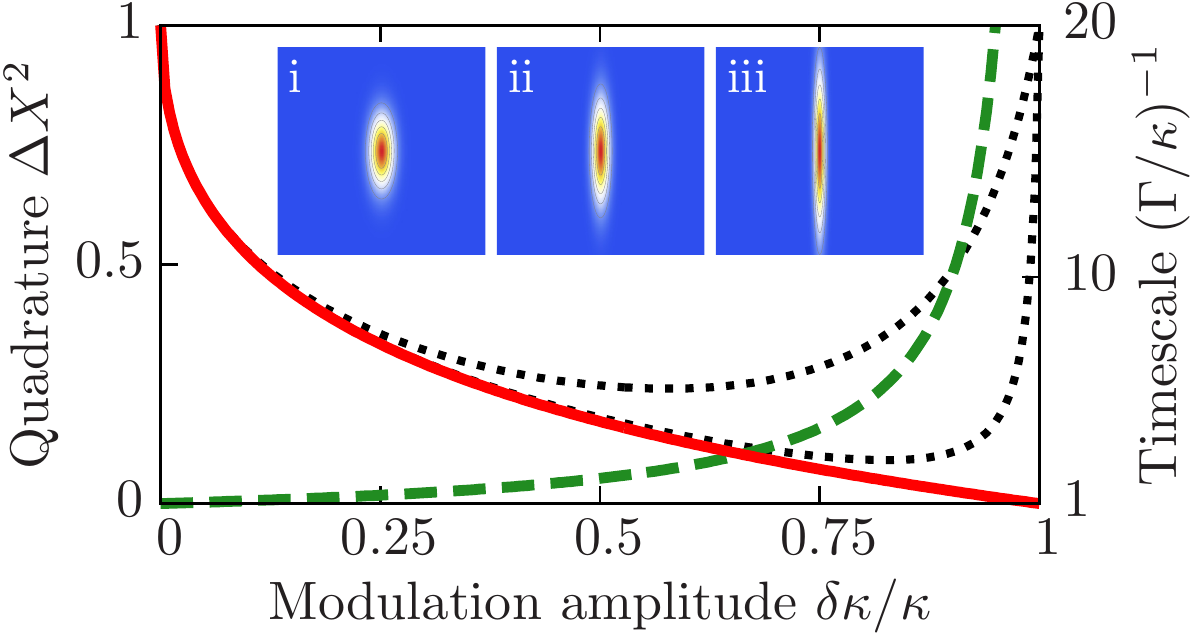}
\caption{(Color online) Squeezed quadrature $\Delta X^2$ (full and dotted lines) and timescale $(\Gamma/\kappa)^{-1}$ (dashed line) as a function of the damping modulation amplitude $\delta\kappa$.
When the environmental contribution at $3\omega_r$ is eliminated (red line), 
as the modulation amplitude increases the squeezed quadrature is reduced and the timescale to reach the steady state increases.
The effect of the bath spectrum at $3\omega_r$ is to limit the minimal squeezing (dotted lines; $\kappa(3\omega_r)/\kappa_c=0.1$ for the leftmost and $0.01$ for the rightmost).
The insets show the Wigner function of the cavity field for $\delta\kappa/\kappa=$0.25 (i), 0.5 (ii), and 0.75 (iii).}
\label{figsqueeze}
\end{figure}

\textit{Monochromatic modulation} -- 
The generation of squeezing by damping is most efficient for a monochromatic modulation at $\omega_m=2\omega_r$ corresponding to $\lambda(t)=\lambda_1\Exp{2i\omega_rt}$.
In the interaction picture, the Lindbladian Eq.~\eqref{genL} then becomes
\begin{equation}
\mathcal{L}=\Gamma\left\{
(\bar{n}+1)\mathcal{D}[\hat{a}]+\bar{n} \mathcal{D}[\hat{a}^\dag]+m \mathcal{S}[\hat{a}]+m^* \mathcal{S}[\hat{a}^\dag]
\right\},
\end{equation}
where we have defined the modulation amplitude $\delta\kappa$, the rate $\Gamma$, the effective thermal number $\bar{n}$,  and the parameter $m$ as
\begin{align}
\delta\kappa&=|\lambda_1|^2\kappa,&
\Gamma&=\kappa-\delta\kappa,&
\bar{n}&=\delta\kappa/\Gamma,&
m&=\lambda_1\kappa/\Gamma.
\end{align}

Under Eq.~\eqref{genL}, a Gaussian state remains Gaussian and is completely specified by the two moments 
\begin{subequations}
\begin{align}
\moy{\hat{a}^\dag\hat{a}}(t)&=\phantom{-}[1-\Exp{-\Gamma t}]\bar{n},\\
\moy{\hat{a}^2}(t)&=-[1-\Exp{-\Gamma t}]m^*,
\end{align}
\end{subequations}
and $\moy{\hat{a}}=0$ without driving. The steady state is reached at the reduced rate $\Gamma<\kappa$. The internal resonator state is then in the vacuum squeezed state $\ket{\xi=r\Exp{i\theta}}$, where the squeezing parameter $\xi$ is fixed by the modulation amplitude and phase: $r=\mathrm{arctanh}|\lambda_1|$ and $\theta=-\arg\lambda_1$.
The variance of the two internal mode quadratures $\hat{X}=\hat{a}^\dag\Exp{i\theta/2}+\hat{a}\Exp{-i\theta/2}$ and $\hat{Y}=i\hat{a}^\dag\Exp{i\theta/2}-i\hat{a}\Exp{-i\theta/2}$ is 
\begin{align}
\Delta X^2&=\frac{1-|\lambda_1|}{1+|\lambda_1|},&
\Delta Y^2&=\frac{1+|\lambda_1|}{1-|\lambda_1|},
\label{quadraturemono}
\end{align}
showing that the resonator state is ideally squeezed with $\Delta X^2\Delta Y^2=1$.
The width of the squeezed quadrature is reduced by increasing the modulation amplitude and vanishes for $|\lambda_1|$ approaching unity.  This is illustrated in Fig.~\ref{figsqueeze}, where the variance $\Delta X^2$ is plotted as a function of the relative modulation amplitude $\delta\kappa/\kappa$ (full red line). The black dotted lines correspond to the result in the absence of a Purcell filter (see the caption), while the green dashed line shows the relevant timescale $1/\Gamma$ for squeezing. While the latter diverges with increasing modulation amplitude, it is possible to beat the 3~dB limit of squeezing at $\delta\kappa\sim0.1\kappa$ with only a slightly reduced rate $\Gamma \sim 0.9 \kappa$.

Following input-output theory~\cite{collett:1984a}, a calculation of the resonator output field is presented in the Appendix~\ref{AppB}. There, we show that while the modulation is active, the relation between the input and output fields is simply a frequency dependent phase shift $\varphi(\omega)=2\arctan[2(\omega-\omega_r)/\Gamma]$. This is the result obtained for an empty cavity of damping rate $\Gamma$. 
However, once the resonator is in the vacuum squeezed state and the modulation is stopped, the resonator is coupled to an environment at zero temperature. The squeezed state then leaks out of the resonator at the original rate $\kappa$ and can be measured by homodyne detection.

\textit{Qubit state-dependent squeezing} -- 
As a natural extension of this squeezing mechanism, we now consider dispersively coupling a qubit to the damping modulated resonator. In addition to producing interesting states of light, this could be used to improve qubit readout in circuit QED. A catch and release protocol for qubit readout based on the setup of Ref.~\cite{yin:2013a} was already theoretically studied by Sete and co-authors~\cite{sete:2013a}.

In the dispersive regime, where the qubit-resonator detuning is much larger than their coupling strength, the qubit-resonator interaction is well approximated by the Hamiltonian
$H_\mathrm{disp}=\chi\hat{\sigma}_z\hat{a}^\dag\hat{a}$, where $\sigma_z$ is the Pauli matrix in the qubit computational basis $\{\ket{0},\ket{1}\}$ and $\chi$ is the dispersive coupling~\cite{blais:2004a}. Here, we focus on the strong dispersive regime where $\chi>\kappa$~\cite{schuster:2007a}. Without damping modulation, the state of the qubit can be measured by driving the resonator such as to displace the resonator field if, for example, the qubit is in state $\ket{1}$ leaving the resonator in the vacuum state if the qubit is in state $\ket{0}$. This is realized by a coherent tone of amplitude $\epsilon$ and frequency $\omega_d=\omega_r+\chi$ driving the resonator. Increasing the drive amplitude helps in further separating the two qubit-state dependent field states thereby improving the measurement fidelity. In practice, from $\sim 1$ to 30 measurement photons are typically used~\cite{vijay:2011a}.

\begin{figure}
\centering
\includegraphics[width=\columnwidth]{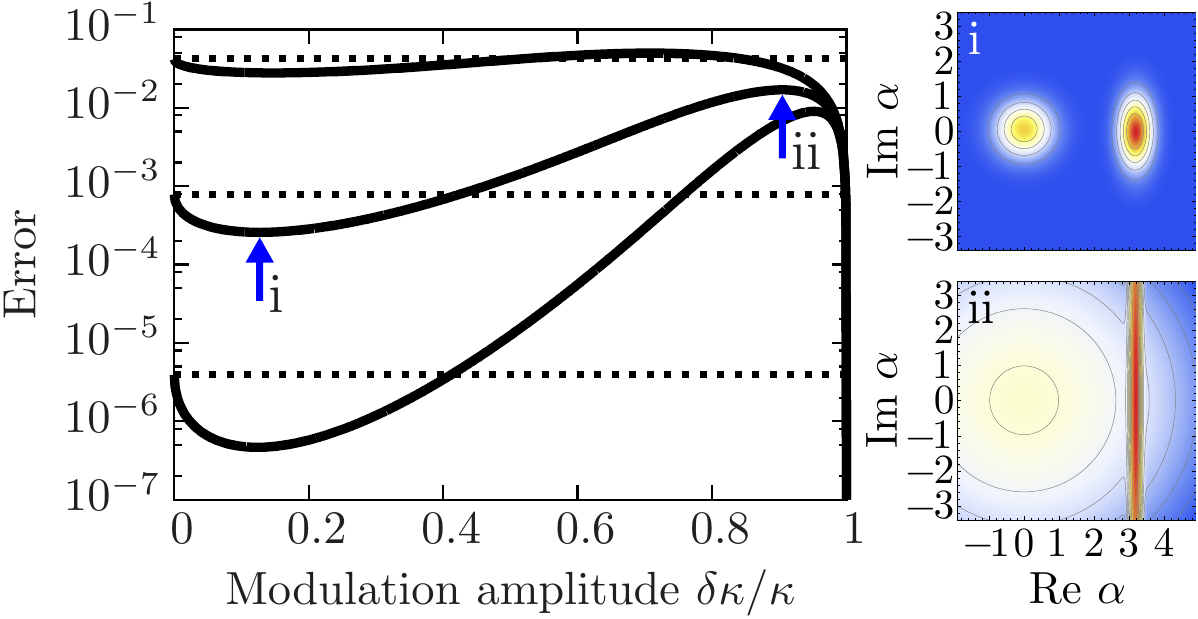}
\caption{(Color online) Measurement error as a function of the modulation amplitude for $\lambda(t)=\sqrt{\delta\kappa}\,\Exp{2i(\omega_r+\chi)t}$. A drive is displacing the field for the state $\ket{1}$ by three (top), ten (middle), twenty (bottom) photons.
The cavity pull is $\chi=10\kappa$.
The dashed line is the error without modulation.
For a displacement of ten photons, the error is minimal at $\delta\kappa=0.13\kappa$ and reduced by a factor of 3.1.
The right panels are the Wigner functions for the minimal error $\delta\kappa/\kappa=0.13$ (i) and for the maximal error $\delta\kappa/\kappa=0.9$ (ii) as indicated by the arrows.}
\label{figerror1}
\end{figure}

To help in further distinguishing the two field states, we propose to modulate the resonator damping such as to produce a squeezed state conditional on the state of the qubit. As before, in the presence of modulation and of the coherent drive, the cavity field is displaced to $\moy{\hat{a}}_1=2\epsilon/\Gamma$ if the qubit is in the excited state. It otherwise remains in the vacuum state, $\moy{\hat{a}}_0=0$. Unless otherwise mentioned, in all numerical calculations we adjust $\epsilon$ such as to have $|\moy{\hat{a}}_1|^2=10$ photons in the resonator. This adjustment depends on the modulation amplitude $\delta \kappa$ via the effective rate $\Gamma$. To squeeze the field state corresponding to the qubit state $\ket{1}$, damping is now modulated at the frequency $\omega_m=2(\omega_r+\chi)$ corresponding to twice the pulled cavity frequency for the qubit in state $\ket 1$, $\omega_r^1 = \omega_r + \chi$.
This modulation generates $\bar{n}$ thermal photons irrespective of the qubit state and squeezes the field corresponding to state $\ket{1}$ to $\moy{\hat{a}^2}_1=-\lambda_1^*\kappa/\Gamma$. On the other hand, if the qubit is in state $\ket 0$, the cavity frequency is pulled to $\omega_r^0 = \omega_r - \chi$. Since $|\omega_m -2\omega_r^0| = 4|\chi| > \kappa$ in the strong dispersive limit, squeezing is negligible. As a result, while the field associated with $\ket 1$ is squeezed, the field associated with $\ket 0$ goes to a thermal state characterized by $\bar n$ thermal photons.

The effect of the drive and the modulation on the resonator's Wigner functions is presented in Fig.~\ref{figerror1}.
As the modulation amplitude $\delta\kappa$ is increased, the field associated with $\ket 1$ is getting more squeezed along the real axis while the resonator state corresponding to $\ket 0$ spreads in all directions. If the goal is to improve qubit readout, there is an optimal value of the modulation amplitude where the overlap between the two Wigner functions is minimal. To quantify the state discrimination, we compute the measurement error
$
E=\frac{1}{2}\int\mathrm{d}x\,\mathrm{Min}[P_0(x),P_1(x)]
$~\cite{sete:2013a}. In this expression, $P_{i}$ is the marginal of the Wigner function corresponding to $\ket{i}$ and integrated along the imaginary axis (see Appendix~\ref{AppE}). 
As seen in Fig.~\ref{figerror1}, the minimal measurement error is obtained for a relatively small modulation amplitude $\delta\kappa=0.13\kappa$. At this point, the error is reduced by a factor of 3 with respect to the error obtained without modulation and for the same field displacement. Because of the small modulation amplitude, the squeezing rate is a large fraction of the bare cavity decay rate with $\Gamma = 0.87 \kappa$. These analytical results based on the dispersive Hamiltonian are in agreement with numerical simulations using the full Jaynes-Cummings coupling.

\begin{figure}
\centering
\includegraphics[width=\columnwidth]{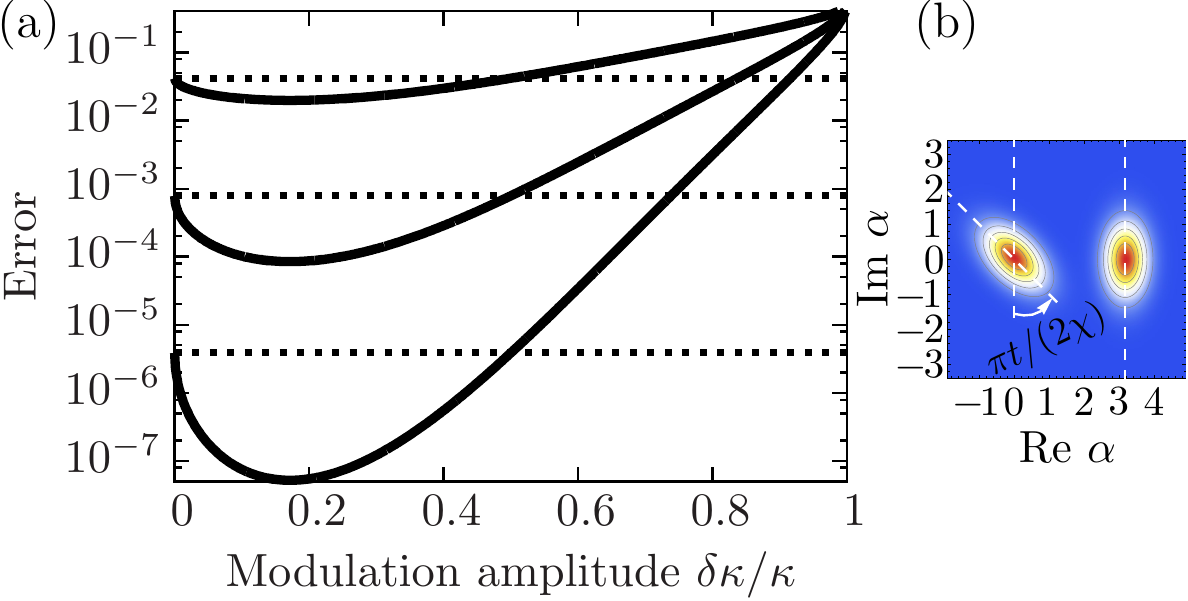}
\caption{(Color online) (a) Measurement error as a function of the modulation amplitude for $\lambda(t)=\sqrt{\delta\kappa/2}\,[\Exp{2i(\omega_r+\chi)t}+\Exp{2i(\omega_r-\chi)t}]$ at the optimal time. A drive is displacing the field for the state $\ket{1}$ by three (top), ten (middle), twenty (bottom) photons.
The cavity pull is $\chi=10\kappa$.
The dashed line is the error without modulation.
For a displacement of ten photons, the error is minimal at $\delta\kappa=0.17\kappa$ and reduced by a factor of 9.3.
(b) Representation of the time evolution of the Wigner functions.
In the rotating frame of the qubit state $\ket{1}$, the Wigner function of the resonator with the qubit in state $\ket{0}$ oscillates at the frequency $4\chi$.}
\label{figerror2}
\end{figure}

\textit{Bichromatic modulation} -- 
Interestingly, it is possible to simultaneously squeeze the two field states corresponding to the two qubit states by applying a modulation oscillating at the qubit-state-dependent pulled resonator frequencies:
\begin{equation}
\lambda(t)=\lambda_0\Exp{2i(\omega_r-\chi)t}+\lambda_1\Exp{2i(\omega_r+\chi)t}.
\end{equation}
As illustrated in Fig.~\ref{figerror2}b), in this situation after a transient regime the squeezing parameters corresponding to the two field states oscillate at the frequency difference $4\chi$ between the two modulations (see Appendix~\ref{AppC}).
In a frame rotating at $\omega_r+\chi$ and for $\chi\gg\kappa$, the steady state is well approximated by $\moy{\hat{a}^\dag\hat{a}}_{0,1}=\delta\kappa/\Gamma$ and
\begin{align}
\moy{\hat{a}^2}_0&\simeq-m_0^*\Exp{4i\chi t},&
\moy{\hat{a}^2}_1&\simeq-m_1^*,
\end{align}
where now $\delta\kappa=(|\lambda_0|^2+|\lambda_1|^2)\kappa$, $m_{0,1}=\lambda_{0,1}\kappa/\Gamma$, and again $\Gamma=\kappa-\delta\kappa$.
The effect of the modulation is maximal for $\lambda_0=\lambda_1\equiv\lambda/\sqrt{2}$.
The resonator is in a thermal squeezed state~\cite{kim:1989a} with quadratures 
$\Delta X^2\Delta Y^2=(1+\lambda^2+\sqrt{2}\lambda)/(1+\lambda^2-\sqrt{2}\lambda)$.
Squeezing is not ideal, 
$\Delta X^2\Delta Y^2=(\kappa^2+\delta\kappa^2)/\Gamma^2$, 
and limited by $\Delta X^2\geq1/\sqrt{2}$ obtained at $\delta\kappa/\kappa=3-2\sqrt{2}$.
This is due to the fact that both modulation frequencies contribute to the thermal photon number $\bar{n}$, but only the resonant modulation contributes to squeeze a particular field state.

With a drive of amplitude $\epsilon$ and frequency $\omega_d=\omega_r+\chi$, the error oscillates in time at the frequency $4\chi$ and reaches the minimal value
\begin{equation}
E_\mathrm{min}=\frac{1}{2}-\frac{1}{2}\mathrm{erf}\left[\frac{2\epsilon}{\sqrt{2\Gamma(\kappa+\delta\kappa-\sqrt{2\kappa\delta\kappa})}}\right].
\label{errorbi}
\end{equation}
This expression is plotted in Fig.~\ref{figerror2}a). The error is decreased by a factor of 9 with respect to the result without modulation and the same field displacement. The minimal error is obtained when the variance of each state is minimal. Experimentally, at the price of losing measurement photons, a stroboscopic observation following the $4\chi$-periodic evolution of the quadrature is required.

\textit{Conclusion} -- 
We have shown that modulating the damping rate of a resonator at twice its natural  frequency creates a vacuum squeezed state.
This approach does not require an optical non-linearity and has no limitation on the squeezed quadrature for the intra-resonator field. This could be realized with existing circuit QED setups. Moreover, by coupling a qubit to the resonator, it is possible to obtain qubit state-dependent squeezing. This could be used to improve the qubit readout fidelity with only a modest modulation amplitude. Quantum bath engineering with a dynamical damping is thus a powerful resource for the generation of non-classical states of light.

\textit{Acknowledgments} -- 
We thank C.~M\"uller, J. Bourassa, A.~A.~Clerk, A.~Wallraff, and J.~M.~Martinis for discussions and acknowledge financial support by CIFAR and NSERC.
After this work was completed, we became aware of an alternate and distinct proposal for dissipative squeezing by A.~Kronwald, F.~Marquardt and A.~A.~Clerk~\cite{clerk2013}.

\appendix

\section{Derivation of the Lindbladian}
\label{AppA}

We consider a single mode of a linear resonator described by the Hamiltonian $H_S=\omega_r \hat{a}^\dag \hat{a}$.
The system is linearly coupled to a bath at zero temperature. The environment is described by the creation operator $\hat{f}^\dag(\omega)$ and characterized by the density of states $d(\omega)$ and the coupling strength $u(\omega)$.
We assume that the system--bath coupling $u(\omega)$ is modulated in time.  It is convenient to represent this by writing 
\begin{equation}\label{eq:Lambda}
\Lambda(t)=1+\lambda(t),
\end{equation}
with the Fourier decomposition $\lambda(t)=\sum_n\lambda_n\,\Exp{i\omega_nt}$ normalized such that the constant term $n=0$ is zero.
The system-bath Hamiltonian reads
\begin{align}
H_\mathrm{int}
&=\int_0^\infty\mathrm{d}\omega\,\sqrt{d(\omega)} \left[u^*(\omega)\Lambda^*(t)\hat{f}^\dag(\omega)\right.\nonumber\\
&\hspace{7.5mm}\left.+u(\omega)\Lambda(t)\hat{f}(\omega)\right] \left[\hat{a}^\dag+\hat{a}\right]\nonumber\\
&\equiv\left[\Lambda^*(t)\hat{F}^\dag+\Lambda(t)\hat{F}\right]\left[\hat{a}^\dag+\hat{a}\right],
\label{eq:Hint}
\end{align}
and the bath Hamiltonian is
\begin{equation}
H_B=\int_0^\infty\mathrm{d}\omega\,\omega\, \hat{f}^\dag(\omega)\hat{f}(\omega).
\end{equation}
The first term on the right-hand-side of Eq.~\eqref{eq:Lambda} leads to the standard decay of the resonator field at a rate which we will denote $\kappa$.

We use here the usual Born-Markov approximations to derive a master equation for the reduced resonator state $\rho(t) = \mathrm{Tr}_B \rho_{\mathrm{tot}}(t)$ after having traced out the environmental degrees of freedom~\cite{gardiner:2004b}.  As usual, we first move to the interaction picture where the interaction Hamiltonian reads
\begin{equation}
\tilde{H}=\left[\Lambda^*(t)\hat{\tilde{F}}^\dag(t)+\Lambda(t)\hat{\tilde{F}}(t)\right]
\left[\hat{a}^\dag\,\Exp{i\omega_rt}+\hat{a}\,\Exp{-i\omega_rt}\right],
\end{equation}
with $\hat{\tilde{F}} = \hat{F} e^{-i\omega t}$. In this frame, the dynamics of $\rho(t)$ is described by the equation
\begin{multline}
\dot{\tilde{\rho}}(t)
=\mathrm{Tr}_B\left\{\int_0^t\mathrm{d}t'\Big( \tilde{H}(t)\rho_{\mathrm{tot}}(t')\tilde{H}(t')\right.\\\left.-\tilde{H}(t)\tilde{H}(t')\rho_{\mathrm{tot}}(t') \Big)\right\}\ + \mathrm{H.c.}
\end{multline}
Assuming that the system-environment state is initially factorized and that the coupling to the environment is weak, we write $\rho_{\mathrm{tot}} (t') = \rho(t')\otimes\rho_B (0)$ in the above expression. Having taken the environment to be at zero temperature, only the terms of the form $\hat{F}^\dag\rho_{\mathrm{tot}}\hat{F}$ and $\hat{F}\hat{F}^\dag\rho_{\mathrm{tot}}$ survive and we arrive at
\begin{align}
\dot{\tilde{\rho}}(t)
=&\int_0^t\mathrm{d}t' \, \Lambda^*(t)\Lambda(t')\,
\moy{\hat{\tilde{F}}(t')\hat{\tilde{F}}^\dag(t)}_B\,
\left( \hat{a}^\dag\,\Exp{i\omega_rt} +\hat{a}\,\Exp{-i\omega_rt} \right) \nonumber\\&\times \tilde{\rho}
\left( \hat{a}^\dag\,\Exp{i\omega_rt'}+\hat{a}\,\Exp{-i\omega_rt'}\right)\nonumber\\
-&\int_0^t\mathrm{d}t' \, \Lambda^*(t')\Lambda(t)\,
\moy{\hat{\tilde{F}}(t)\hat{\tilde{F}}^\dag(t')}_B\,
\left( \hat{a}^\dag\,\Exp{i\omega_rt} +\hat{a}\,\Exp{-i\omega_rt} \right) \nonumber\\&\times
\left( \hat{a}^\dag\,\Exp{i\omega_rt'}+\hat{a}\,\Exp{-i\omega_rt'}\right)\tilde{\rho}\ + \mathrm{H.c.}
\end{align}
Given that $\moy{\hat{f}(\omega)\hat{f}^\dag(\omega')}_B=\delta(\omega-\omega')$,
we have 
$\displaystyle \moy{\hat{\tilde{F}}(t)\hat{\tilde{F}}^\dag(t')}_B=\int_0^\infty\mathrm{d}\omega\,\Exp{-i\omega(t-t')} \kappa(\omega)$ where $\kappa(\omega)\propto d(\omega)|u(\omega)|^2$ such that
\begin{align}
\dot{\tilde{\rho}}(t)=&\sum_{n,m}\Lambda_n^*\Lambda_m\,\Exp{i(\omega_m-\omega_n)t}\int_0^t\mathrm{d}\tau\int_0^\infty\mathrm{d}\omega\nonumber\\
\times\Big\{&\kappa(\omega)\,\Exp{\phantom{-}i(\omega-\omega_m)\tau}
\left( \hat{a}^\dag\,\Exp{i\omega_rt} +\hat{a}\,\Exp{-i\omega_rt} \right) \nonumber\\&\times \tilde{\rho}
\left( \hat{a}^\dag\,\Exp{i\omega_r(t-\tau)}+\hat{a}\,\Exp{-i\omega_r(t-\tau)}\right)\nonumber\\
-&\kappa(\omega)\,\Exp{-i(\omega-\omega_n)\tau}
\left( \hat{a}^\dag\,\Exp{i\omega_rt} +\hat{a}\,\Exp{-i\omega_rt} \right) \nonumber\\&\times 
\left( \hat{a}^\dag\,\Exp{i\omega_r(t-\tau)}+\hat{a}\,\Exp{-i\omega_r(t-\tau)}\right)\tilde{\rho}\quad\Big\}
+\mathrm{H.c.}
\end{align}
where $\tau=t-t'$ and we set $\Lambda_{n\neq0}=\lambda_n$ and $\Lambda_0=1$.
At long times $t\to\infty$ and neglecting principal parts, the dynamics is finally given by
\begin{align}
\dot{\tilde{\rho}}(t)&=\frac{1}{2}\sum_{n,m}\Lambda_n^*\Lambda_m\,\Exp{i(\omega_m-\omega_n)t}\nonumber\\
\times\Big\{&\kappa(\omega_n+\omega_r)\left[
\hat{a}\tilde{\rho}\hat{a}^\dag-\hat{a}^\dag\hat{a}\tilde{\rho}
+(\hat{a}\tilde{\rho}\hat{a}-\hat{a}\hat{a}\tilde{\rho})\,\Exp{-2i\omega_rt}\right]\nonumber\\
+&\kappa(\omega_n-\omega_r)\left[
\hat{a}^\dag\tilde{\rho}\hat{a}-\hat{a}\hat{a}^\dag\tilde{\rho}
+(\hat{a}^\dag\tilde{\rho}\hat{a}^\dag-\hat{a}^\dag\hat{a}^\dag\tilde{\rho})\,\Exp{2i\omega_rt}\right]
\Big\}\nonumber\\
+&\mathrm{H.c.}
\end{align}
Moving back to the laboratory frame, we then find
\begin{align}
\dot{\rho}(t)
=&-i[\omega_r \hat a^\dag \hat a,\rho]
+\frac{1}{2}\sum_{n,m}\Lambda_n^*\Lambda_m\Exp{i(\omega_m-\omega_n)t}\nonumber\\
\times\Big\{&[\kappa(\omega_m+\omega_r)+\kappa(\omega_n+\omega_r)]\hat{a}\rho\hat{a}^\dag\nonumber\\
&-\kappa(\omega_n+\omega_r)\hat{a}^\dag \hat{a}\rho-\kappa(\omega_m+\omega_r)\rho \hat{a}^\dag \hat{a}\nonumber\\
+&[\kappa(\omega_m-\omega_r)+\kappa(\omega_n-\omega_r)]\hat{a}^\dag\rho\hat{a}\nonumber\\
&-\kappa(\omega_n-\omega_r)\hat{a}\hat{a}^\dag \rho-\kappa(\omega_m-\omega_r)\rho\hat{a} \hat{a}^\dag\nonumber\\
+&[\kappa(\omega_m-\omega_r)+\kappa(\omega_n+\omega_r)]\hat{a}\rho\hat{a}\nonumber\\
&-\kappa(\omega_n+\omega_r)\hat{a}\hat{a}\rho-\kappa(\omega_m-\omega_r)\rho \hat{a}\hat{a}\nonumber\\
+&[\kappa(\omega_m+\omega_r)+\kappa(\omega_n-\omega_r)]\hat{a}^\dag\rho\hat{a}^\dag\nonumber\\
&-\kappa(\omega_n-\omega_r)\hat{a}^\dag \hat{a}^\dag\rho-\kappa(\omega_m+\omega_r)\rho \hat{a}^\dag \hat{a}^\dag\Big\}.
\end{align}

As discussed in the main text, the terms proportional to $\hat a^2$ and $\hat a^{\dag2}$ can be dropped by invoking the rotating-wave approximation unless the modulation cancels their time dependence. This occurs for a modulation at $2\omega_r$. Therefore, assuming that $\omega_n \sim 2\omega_r$ and taking into account that $\kappa(\omega <Ê0) = 0$ since the bath is at zero temperature, we find that the resonator master equation takes the form $\dot{\rho}=-i[\omega_r\hat{a}^\dag\hat{a},\rho]+\mathcal{L}\rho$, where the Lindbladian $\mathcal{L}\cdot$ is
\begin{multline}\label{eq:SMgenL}
\mathcal{L}=(1+\beta|\lambda(t)|^2)\kappa \mathcal{D}[\hat{a}]+|\lambda(t)|^2\kappa \mathcal{D}[\hat{a}^\dag]\\
+\lambda(t)\kappa \mathcal{S}[\hat{a}]+\lambda^*(t)\kappa \mathcal{S}[\hat{a}^\dag],
\end{multline}
where $\kappa=\kappa(\omega_r)\simeq\kappa(\omega_n-\omega_r)$, $\beta=\kappa(3\omega_r)/\kappa$, and
\begin{align}
\mathcal{D}[\hat{a}]\cdot &=\hat{a} \cdot \hat{a}^\dag - \tfrac{1}{2}\{\hat{a}^\dag \hat{a},\cdot\},\\
\mathcal{S}[\hat{a}]\cdot &=\hat{a} \cdot \hat{a}- \tfrac{1}{2}\{\hat{a}^2,\cdot\}.
\end{align}
The corresponding squeezed quadrature is minimal at $\delta\kappa=\kappa/(1+\sqrt{\beta})^2$ and equal to $\sqrt{\beta}/(1+\sqrt{\beta})$.
If we assume that the resonator's environment is ohmic, $\beta=3$. This contribution can be eliminated with a Purcell filter~\cite{reed:2010a}.

\section{Monochromatic modulation}
\label{AppB}

We focus here on the situation where the modulation is monochromatic with $\lambda(t)=\lambda_1\Exp{2i\omega_rt}$.

\subsection{Squeezing transformation}

As discussed in the main text, the above Lindbladian does not map back to the standard degenerate parametric amplifier. This can be verified by applying a squeezing transformation on the Lindbladian. As will be clear below, this also illustrates how damping is responsible for squeezing. 

We start with the master equation derived in the previous section, which in the presence of a monochromatic modulation and in the laboratory frame reads 
\begin{multline}
\dot \rho = -i[\omega_r \hat{a}^\dag \hat{a},\rho] +
\kappa \mathcal{D}[\hat{a}]\rho+|\lambda_1|^2 \kappa \mathcal{D}[\hat{a}^\dag]\rho\\
+\lambda_1\Exp{2i\omega_rt} \kappa \mathcal{S}[\hat{a}]\rho+\lambda_1^*\Exp{-2i\omega_rt} \kappa \mathcal{S}[\hat{a}^\dag]\rho.
\end{multline}
Moving to the interaction picture we therefore have
\begin{equation}
\begin{split}
\dot {\tilde\rho} 
&
=\kappa \mathcal{D}[\hat{a}]\tilde\rho+|\lambda_1|^2 \kappa \mathcal{D}[\hat{a}^\dag]\tilde\rho +\lambda_1 \kappa \mathcal{S}[\hat{a}]\tilde\rho\\
&+\lambda_1^* \kappa \mathcal{S}[\hat{a}^\dag]\tilde\rho.
\end{split}
\end{equation}

We now go to a squeezed frame by defining $\tilde\rho'  = S^\dag \tilde\rho S$, where
\begin{equation}
S=\exp\left\{\frac{1}{2}\xi^*\hat{a}^2-\frac{1}{2}\xi\hat{a}^{\dag2}\right\},
\end{equation}
is the squeezing operator~\cite{gardiner:2004b} and where we take $\xi=r\Exp{i\theta}$. The master equation for $\tilde\rho'$ reads
\begin{multline}
\dot {\tilde\rho}' 
=\kappa \mathcal{D}[S^\dag \hat{a} S]\tilde\rho'+|\lambda_1|^2 \kappa \mathcal{D}[S^\dag  \hat{a}^\dag S]\tilde\rho'\\
+\lambda_1 \kappa \mathcal{S}[S^\dag  \hat{a} S]\tilde\rho'+\lambda_1^* \kappa \mathcal{S}[S^\dag \hat{a}^\dag S]\tilde\rho'.
\end{multline}

Using $S^\dag \hat aS=\cosh r \,\hat a-\Exp{-i\theta}\sinh r \,\hat a^\dag$, we find
\begin{align}
\mathcal{D}[S^\dag \hat{a} S] 
&= \cosh^2r\,\mathcal{D}[\hat a] +\sinh^2r\,\mathcal{D}[\hat a^\dag] \nonumber\\
-&\tfrac{1}{2}\sinh2r\,\Exp{-i\theta}\mathcal{S}[\hat a] -\tfrac{1}{2}\sinh2r\,\Exp{i\theta}\mathcal{S}[\hat a^\dag] ,\\
\mathcal{D}[S^\dag \hat{a}^\dag S] 
&= \sinh^2r\,\mathcal{D}[\hat a] +\cosh^2r\,\mathcal{D}[\hat a^\dag] \nonumber\\
-&\tfrac{1}{2}\sinh2r\,\Exp{-i\theta}\mathcal{S}[\hat a] -\tfrac{1}{2}\sinh2r\,\Exp{i\theta}\mathcal{S}[\hat a^\dag] ,\\
\mathcal{S}[S^\dag \hat{a} S] 
&= \big(-\tfrac{1}{2}\sinh2r\,\mathcal{D}[\hat a] -\tfrac{1}{2}\sinh2r\,\mathcal{D}[\hat a^\dag] \nonumber\\
+&\cosh^2r\,\Exp{-i\theta}\mathcal{S}[\hat a] +\sinh^2r\,\Exp{i\theta}\mathcal{S}[\hat a^\dag] \big)\Exp{i\theta},\\
\mathcal{S}[S^\dag \hat{a}^\dag S] 
&= \big(-\tfrac{1}{2}\sinh2r\,\mathcal{D}[\hat a] -\tfrac{1}{2}\sinh2r\,\mathcal{D}[\hat a]^\dag \nonumber\\
+&\sinh^2r\,\Exp{-i\theta}\mathcal{S}[\hat a] +\cosh^2r\,\Exp{i\theta}\mathcal{S}[\hat a^\dag] \big)\Exp{-i\theta}.
\end{align}
Putting everything together with the choice $\theta = -\arg\lambda_1$ and $r=\mathrm{arctanh}|\lambda_1|$,
the transformed master equation takes the simple form
\begin{equation}
\dot {\tilde\rho}' = \Gamma\mathcal{D}[\hat{a}] \tilde\rho'.
\end{equation}
Going back to a frame that is not rotating at the resonator frequency, we finally get 
\begin{equation}
\dot {\rho}' =  -i[\omega_r \hat a^\dag \hat a ,\rho'] + \Gamma\mathcal{D}[\hat{a}] \rho'.
\end{equation}
We thus find that modulating the damping rate is equivalent to cooling the transformed linear oscillator $\rho'$ to its ground state at the renormalized rate $\Gamma$. Because of the squeezing transformation, the ground state $\ket{0}$ of $\rho'$ corresponds to the ideally squeezed state $\ket{\xi}$ in the laboratory frame. In short, a zero temperature bath is damping the system to an ideally squeezed state.  We also emphasize that no nonlinear term is obtained in the transformed Hamiltonian or Lindbladian, showing that the dissipative and coherent squeezing are different mechanisms with different bounds.

\subsection{Input-ouput theory}

We use the standard input-output formalism~\cite{collett:1984a} to compute the output field in the presence of damping modulations. We take $\lambda_1$ to be real to simplify the calculation. Using the system-bath Hamiltonian in the interaction picture, the equation of motion of the cavity field is
\begin{multline}
\dot{\hat{a}}=-i\int_0^\infty\mathrm{d}\omega \sqrt{d(\omega)}
\left[u(\omega)\Lambda(t)\,\Exp{-i\omega t}\hat{f}(\omega)\right.\\\left.+u^*(\omega)\Lambda^*(t)\,\Exp{i\omega t}\hat{f}^\dag(\omega)\right]\Exp{i\omega_r t},
\end{multline}
while for the environment we have
\begin{equation}\label{eq:dotf}
\dot{\hat{f}}(\omega)=-i\sqrt{d(\omega)}u^*(\omega)\Lambda^*(t)\left[\Exp{i(\omega+\omega_r)t}\hat{a}^\dag
+
\Exp{i(\omega-\omega_r)t}\hat{a}
\right].
\end{equation}
Integrating the previous expression from an initial condition $\hat{f}(\omega,t_0)$ at time $t_0<t$ yields
\begin{multline}
\hat{f}(\omega,t)=\hat{f}(\omega,t_0)\\-i\sqrt{d(\omega)}u^*(\omega)\int_{t_0}^t\mathrm{d}t'\Lambda^*(t')
\left[\Exp{i(\omega+\omega_r)t'}\hat{a}^\dag(t')
\right.\\\left.+
\Exp{i(\omega-\omega_r)t'}\hat{a}(t')
\right].
\end{multline}
Inserting this results in the expression for $\dot{\hat a}$ then leads to
\begin{align}
\dot{\hat{a}}
=&
-i\int_0^\infty\mathrm{d}\omega \sqrt{d(\omega)}u(\omega) \Lambda(t)\,\Exp{-i(\omega-\omega_r) t} \hat{f}(\omega,t_0)\nonumber\\
&-i\int_0^\infty\mathrm{d}\omega \sqrt{d(\omega)}u^*(\omega) \Lambda^*(t)\,\Exp{i(\omega+\omega_r)t} \hat{f}^\dag(\omega,t_0)\nonumber\\
&-\int_0^\infty\mathrm{d}\omega\int_{t_0}^t\mathrm{d}t'd(\omega)|u(\omega)|^2\Lambda(t)\Lambda^*(t')\,\Exp{-i(\omega-\omega_r)t}\nonumber\\&\qquad\times
\left[\Exp{i(\omega+\omega_r)t'}\hat{a}^\dag(t')
+
\Exp{i(\omega-\omega_r)t'}\hat{a}(t')
\right]\nonumber\\
&+\int_0^\infty\mathrm{d}\omega\int_{t_0}^t\mathrm{d}t'd(\omega)|u(\omega)|^2\Lambda^*(t)\Lambda(t')\,\Exp{i(\omega+\omega_r)t}\nonumber\\&\qquad\times
\left[\Exp{-i(\omega-\omega_r)t'}\hat{a}^\dag(t')
+
\Exp{-i(\omega+\omega_r)t'}\hat{a}(t')
\right].
\end{align}
Now, we expand $\Lambda(t)$ and we make the rotating wave approximation by dropping all the fast oscillating terms and then safely take the lower limit of the integral over the frequency to $-\infty$~\cite{gardiner:2004b}. 
Additionally, we define the interaction picture input field as~\cite{collett:1984a}
\begin{equation}
\hat{a}_\mathrm{in}
=
-i\frac{1}{\sqrt{2\pi}}\int_{-\infty}^\infty\mathrm{d}\omega \,\Exp{-i(\omega-\omega_r) t} \hat{f}(\omega,t_0).
\end{equation}

With $\displaystyle\int_{-\infty}^\infty\mathrm{d}\omega\int_{t_0}^t\mathrm{d}t'd(\omega)|u(\omega)|^2\,\Exp{-i(\omega-\omega_0)(t-t')}=\kappa(\omega_0)/2$ we finally obtain the Langevin equation for $\hat a$:
\begin{equation}\label{eq:aLangevinIn}
\dot{\hat{a}}=\sqrt{\kappa}\hat{a}_\mathrm{in}-\lambda_1\sqrt{\kappa}\hat{a}_\mathrm{in}^\dag-\tfrac{1}{2}\Gamma \hat{a}.
\end{equation}
In this expression, $\Gamma=[1-(1+\beta)\lambda_1^2]\kappa$, and in the following we take $\beta=0$, assuming that a Purcell filter is present. Now, integrating the equation of motion, Eq.~\eqref{eq:dotf}, from a final condition $\hat{f}(\omega,t_1)$ at $t_1>t$ and defining
\begin{equation}
\hat{a}_\mathrm{out}=i\frac{1}{\sqrt{2\pi}}\int_{-\infty}^\infty\mathrm{d}\omega\,\Exp{-i(\omega-\omega_r) t} \hat{f}(\omega,t_1),
\end{equation}
we also have for $\dot{\hat a}$
\begin{equation}\label{eq:aLangevinOut}
\dot{\hat{a}}=-\sqrt{\kappa}\hat{a}_\mathrm{out}+\lambda_1\sqrt{\kappa}\hat{a}_\mathrm{out}^\dag+\tfrac{1}{2}\Gamma \hat{a}.
\end{equation}
From the two Langevin equations for $\dot{\hat a}$, we find that the input and output fields are related by
\begin{equation}
\hat{a}_\mathrm{in}+\hat{a}_\mathrm{out}-\lambda_1(\hat{a}_\mathrm{in}^\dag+\hat{a}_\mathrm{out}^\dag)=(1-\lambda_1^2)\sqrt{\kappa}\hat{a}.
\end{equation}

We now define the Fourier transform 
$\displaystyle \hat{a}(\omega)=\frac{1}{\sqrt{2\pi}}\int_{-\infty}^\infty\mathrm{d}t\,\Exp{i\omega t}\hat{a}(t)$ 
and the same for $\hat{a}_\mathrm{in}(\omega)$ and $\hat{a}_\mathrm{out}(\omega)$.
We note $\hat{a}^\dag(\omega)$ the Fourier transform of $a^\dag(t)$.
From Eq.~\eqref{eq:aLangevinIn} we find that the intra-cavity field $\hat{a}$ is linked to the input field $\hat{a}_\mathrm{in}$ through
\begin{align}
\begin{pmatrix}\hat{a}(\omega)\\\hat{a}^\dag(\omega)\end{pmatrix}
&=M(\omega)
\begin{pmatrix}\hat{a}_\mathrm{in}(\omega)\\\hat{a}_\mathrm{in}^\dag(\omega)\end{pmatrix},\\
M(\omega)&=
\frac{\sqrt{\kappa}}{\frac{1}{2}\Gamma-i(\omega-\omega_r)}
\begin{pmatrix}1&-\lambda_1\\-\lambda_1&1\end{pmatrix},
\end{align}
where we have reintroduced the resonator frequency.  The intra-cavity quadratures are calculated using $\moy{\hat{a}_\mathrm{in}(\omega)\hat{a}^\dag_\mathrm{in}(\omega')}=\delta(\omega+\omega')$ and give the same results as with a quantum master equation approach.
On the other hand, from Eq.~\eqref{eq:aLangevinOut} the output field $\hat{a}_\mathrm{out}$ is linked to the intra-cavity field $\hat{a}$ through
\begin{align}
\begin{pmatrix}\hat{a}_\mathrm{out}(\omega)\\\hat{a}_\mathrm{out}^\dag(\omega)\end{pmatrix}
&=M^{-1}(-\omega)
\begin{pmatrix}\hat{a}(\omega)\\\hat{a}^\dag(\omega)\end{pmatrix},\\
M^{-1}(-\omega)&=
\frac{\frac{1}{2}\Gamma+i(\omega-\omega_r)}{\sqrt{\kappa}(1-\lambda_1^2)}
\begin{pmatrix}1&\lambda_1\\\lambda_1&1\end{pmatrix}.
\end{align}
Using these two relations, we find the relation between the input field and output fields in frequency space:
\begin{align}
\begin{pmatrix}\hat{a}_\mathrm{out}(\omega)\\\hat{a}_\mathrm{out}^\dag(\omega)\end{pmatrix}
&=M^{-1}(-\omega)M(\omega)
\begin{pmatrix}\hat{a}_\mathrm{in}(\omega)\\\hat{a}_\mathrm{in}^\dag(\omega)\end{pmatrix}\\
&=\frac{\frac{1}{2}\Gamma+i(\omega-\omega_r)}{\frac{1}{2}\Gamma-i(\omega-\omega_r)}
\begin{pmatrix}\hat{a}_\mathrm{in}(\omega)\\\hat{a}_\mathrm{in}^\dag(\omega)\end{pmatrix}\\
&=\Exp{i\varphi(\omega)}
\begin{pmatrix}\hat{a}_\mathrm{in}(\omega)\\\hat{a}_\mathrm{in}^\dag(\omega)\end{pmatrix},
\end{align}
with $\varphi=2\arctan[2(\omega-\omega_r)/\Gamma]$.
There is a frequency dependent phase shift between the input and the output, similar to what happens for an empty cavity of damping rate $\Gamma$.

\section{Bichromatic modulation}
\label{AppC}

As discussed in the main text, we consider a qubit that is dispersively coupled to the resonator. In this dispersive regime, the qubit-resonator Hamiltonian is well approximated by $H_\mathrm{dips}=(\omega_r+\chi\sigma_z)\hat{a}^\dag\hat{a}$. In effect, the qubit is pulling the resonator frequency to a qubit-state dependent value $\omega_r^1=\omega_r+\chi$ and $\omega_r^0=\omega_r-\chi$, leading to two distinct cavity fields in the presence of a drive. 

\begin{figure}[t]
\centering
\includegraphics[height=4.5cm]{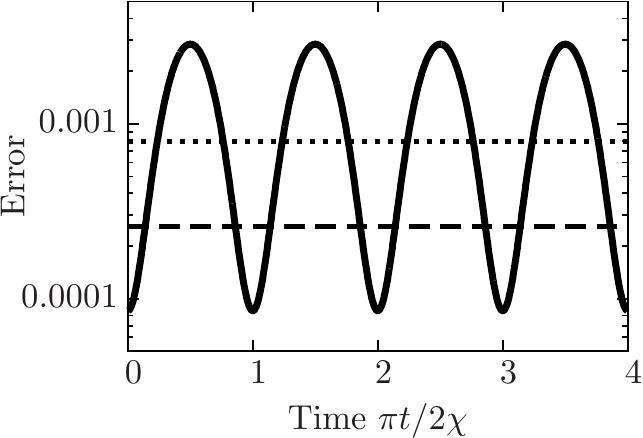}
\caption{Measurement error as a function of time for $\lambda(t)=\sqrt{\delta\kappa/2}\,[\Exp{2i(\omega_r+\chi)t}+\Exp{2i(\omega_r-\chi)t}]$, $\delta\kappa=0.17\kappa$ and a drive displacing the field for the state $\ket{1}$ by ten photons.
The frequency shift is $\chi=10\kappa$.
The dotted line is the error without modulation and the dashed line is the minimal error for a monochromatic modulation.}
\label{figerrort}
\end{figure}

We now imagine modulating the resonator damping such that
\begin{equation}
\lambda(t)=\lambda_0\,\Exp{i2\omega_r^0 t}+\lambda_1\,\Exp{i2\omega_r^1 t}.
\end{equation}
We further take $\lambda_0=\lambda_1=\lambda/\sqrt{2}$ and assume that a Purcell filter is present such that $\beta=0$. In the interaction picture of the bare resonator, the Lindbladian Eq.~\eqref{eq:SMgenL} then reads
\begin{multline}
\mathcal{L}
=\kappa L[\hat{a}]
+\left[1+\cos(4\chi t)\right]^2\lambda^2\kappa L[\hat{a}^\dag]
\\+\sqrt{2}\lambda\cos(2\chi t)\kappa L'[\hat{a}]
+\sqrt{2}\lambda\cos(2\chi t)\kappa L'[\hat{a}^\dag].
\end{multline}
The equations of motion for the two moments completely characterizing the Gaussian state of the resonator are
\begin{align}
\partial_t\moy{\hat{a}^\dag\hat{a}}=&-\{\kappa-[1+\cos(4\chi t)]^2\lambda^2\kappa\}\moy{\hat{a}^\dag\hat{a}}\nonumber\\
&+\left[1+\cos(4\chi t)\right]^2\lambda^2\kappa,\\
\partial_t\moy{\hat{a}^2}=&-\{2i\chi\moy{\sigma_z}+\kappa-[1+\cos(4\chi t)]^2\lambda^2\kappa\}\moy{\hat{a}^2}\nonumber\\
&-\sqrt{2}\cos(2\chi t)\lambda\kappa.
\end{align}
Solving these equations starting from the vacuum, we find that the long time behavior is a Fourier series with the fundamental frequency $4\chi$ and prefactor $I_n[\lambda^2\kappa/(4\chi)]$, with $I_n(x)$ the modified Bessel function. In the limit where $\chi\gg\kappa$, the leading term is obtained for $I_0[\lambda^2\kappa/(4\chi)]\simeq1$.
The temporal evolution is then
\begin{align}
\moy{\hat{a}^\dag\hat{a}}(t)\simeq&\frac{\lambda^2}{1-\lambda^2}\nonumber\\
&+\lambda^2\frac{(1-\lambda^2)\cos(4\chi t)+4(\chi/\kappa)\sin(4\chi t)}{(1-\lambda^2)^2+16(\chi/\kappa)^2}\nonumber\\
\simeq&\frac{\lambda^2}{1-\lambda^2},\\
\moy{\hat{a}^2}(t)\simeq&-\frac{\lambda}{\sqrt{2}}\left[\frac{\Exp{2i\chi t}}{1-\lambda^2+2i(1+\moy{\sigma_z})\chi/\kappa}\right.\nonumber\\&\left.+\frac{\Exp{-2i\chi t}}{1-\lambda^2-2i(1-\moy{\sigma_z})\chi/\kappa}\right]\nonumber\\
\simeq&-\frac{\lambda}{\sqrt{2}}\frac{\Exp{-2i\chi\moy{\sigma_z}t}}{1-\lambda^2}.
\end{align}
The time evolution of the error is plotted in Fig.~\ref{figerrort}.

\section{Modulation of the cavity frequency}
\label{AppD}

We consider a resonator whose frequency is modulated in such a way that
\begin{equation}
H=[\omega_r + \delta\omega_r\cos(\omega_mt)]\hat a^\dag \hat a,
\end{equation}
where $\omega_m$ and $\delta\omega_r$ are the modulation frequency and amplitude, respectively. Going to the interaction picture, the system-environment coupling Hamiltonian reads
\begin{multline}
\tilde{H}=\sum_{n=-\infty}^{\infty}J_n(\delta\omega_r/\omega_m)\left[\hat{\tilde{F}}^\dag(t)+\hat{\tilde{F}}(t)\right]
\\\times\left[\hat{a}^\dag\,\Exp{i(\omega_r+n\omega_m)t}+\hat{a}\,\Exp{-i(\omega_r+n\omega_m)t}\right].
\end{multline}
Following the same procedure as before, we find the master equation for the reduced resonator's state to be
\begin{align}
\dot{\tilde{\rho}}(t)=&\frac{1}{2}\sum_{n,n'}J_n(\delta\omega_r/\omega_m)J_{n'}(\delta\omega_r/\omega_m)\nonumber\\
\times\Big\{&\kappa(\omega_r+n'\omega_m)\hat{a}\tilde{\rho}\hat{a}^\dag\,\Exp{-i(n-n')\omega_mt}\nonumber\\
+&\kappa(-\omega_r-n'\omega_m)\hat{a}^\dag\tilde{\rho}\hat{a}\,\Exp{i(n-n')\omega_mt}\nonumber\\
+&\kappa(\omega_r+n'\omega_m)\hat{a}^\dag\tilde{\rho}\hat{a}^\dag\,\Exp{i(n+n')\omega_mt+2i\omega_rt}\nonumber\\
+&\kappa(-\omega_r-n'\omega_m)\hat{a}\tilde{\rho}\hat{a}\,\Exp{-i(n+n')\omega_mt-2i\omega_rt}\nonumber\\
-&\kappa(\omega_r+n'\omega_m)\hat{a}^\dag \hat{a}\tilde{\rho}\,\Exp{i(n-n')\omega_mt}\nonumber\\
-&\kappa(-\omega_r-n'\omega_m)\hat{a} \hat{a}^\dag\tilde{\rho}\,\Exp{-i(n-n')\omega_mt}\nonumber\\
-&\kappa(-\omega_r-n'\omega_m)\hat{a}^\dag \hat{a}^\dag\tilde{\rho}\,\Exp{i(n+n')\omega_mt+2i\omega_rt}\nonumber\\
-&\kappa(\omega_r+n'\omega_m)\hat{a}\hat{a}\tilde{\rho}\,\Exp{-i(n+n')\omega_mt-2i\omega_rt}\quad\Big\}+\mathrm{H.c.}
\end{align}

\begin{figure}
\centering
\includegraphics[height=4.5cm]{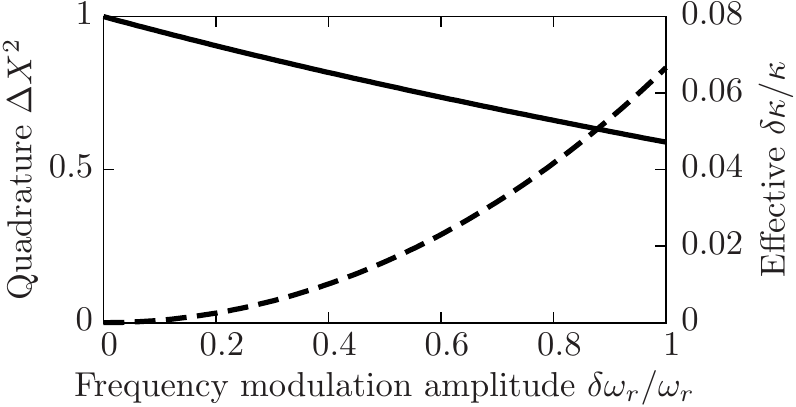}
\caption{Squeezed quadrature (solid line) and effective damping modulation (dashed line) as a function of the amplitude of the frequency modulation for a modulation frequency $\omega_m=2\omega_r$ and keeping only the contribution from terms proportional to $\kappa(\omega_r)$.}
\label{figfreqmod}
\end{figure}

The effect of the modulation is noticeable when the modulation frequency is around $2\omega_r$.
In this situation, $\omega_m=2\omega_r$, the Lindbladian can take the form
\begin{equation}
\mathcal{L}=\Gamma\{(\bar{n}+1) \mathcal{D}[\hat{a}]+\bar{n} \mathcal{D}[\hat{a}^\dag]+m\mathcal{S}[\hat{a}]+m\mathcal{S}[\hat{a}^\dag]\},
\end{equation}
where we have defined
\begin{align}
\Gamma&=\sum_n\kappa[(2n+1)\omega_r]\left[J_n^2(\tfrac{\delta\omega_r}{\omega_m})-J_{n+1}^2(\tfrac{\delta\omega_r}{\omega_m})\right],\\
\bar{n}&=\frac{1}{\Gamma}\sum_n\kappa[(2n+1)\omega_r]J_{n+1}^2(\tfrac{\delta\omega_r}{\omega_m}),\\
m&=\frac{1}{\Gamma}\sum_n(-1)^{n+1}\kappa[(2n+1)\omega_r]J_n(\tfrac{\delta\omega_r}{\omega_m})J_{n+1}(\tfrac{\delta\omega_r}{\omega_m}).
\end{align}
For simplicity, when only the contribution $\kappa\equiv\kappa(\omega_r)$ is taken into account, the quadrature variances take the form
\begin{align}
\Delta X^2&=\frac{J_0(\delta\omega_r/\omega_m)-J_1(\delta\omega_r/\omega_m)}{J_0(\delta\omega_r/\omega_m)+J_1(\delta\omega_r/\omega_m)},\\
\Delta Y^2&=\frac{J_0(\delta\omega_r/\omega_m)+J_1(\delta\omega_r/\omega_m)}{J_0(\delta\omega_r/\omega_m)-J_1(\delta\omega_r/\omega_m)}.
\end{align}
These expressions are equivalent to the case of a damping modulation with the correspondence
\begin{equation}
\frac{\delta\kappa}{\kappa}=\left[\frac{J_1(\delta\omega_r/2\omega_r)}{J_0(\delta\omega_r/2\omega_r)}\right]^2.
\end{equation}
The squeezed quadrature and the effective damping modulation are plotted against the amplitude of the frequency modulation in Fig.~\ref{figfreqmod}. For $\delta\omega_r/\omega_r \sim 0.2~\%$ as in the experimental realization of Ref.~\cite{yin:2013a}, the effect of this contribution is negligible.

\section{Calculation of the error}
\label{AppE}

A Gaussian field is completely determined by $\moy{\hat{a}}$, $\moy{\hat{a}^\dag\hat{a}}$, and $\moy{\hat{a}^2}$.
Let us define the shifted operator $\hat{b}=\hat{a}-\moy{\hat{a}}$ and the angle $\theta=\pi+\arg\moy{\hat{b}^2}$ (such that, for a vacuum squeezed state, $\theta$ corresponds to the phase of the squeezing parameter).
The squeezing direction is given by $\theta/2$ and the squeezed quadratures are $\hat{X}=\hat{a}^\dag\Exp{i\theta/2}+\hat{a}\Exp{-i\theta/2}$ and $\hat{Y}=i\hat{a}^\dag\Exp{i\theta/2}-i\hat{a}\Exp{-i\theta/2}$.
Their variances are
\begin{align}
\Delta X^2&=4\sigma_x^2=2\moy{\hat{b}^\dag\hat{b}}+1-2|\moy{\hat{b}^2}|,\\
\Delta Y^2&=4\sigma_y^2=2\moy{\hat{b}^\dag\hat{b}}+1+2|\moy{\hat{b}^2}|.
\end{align}

For such a Gaussian state, the Wigner function reads
\begin{equation}
W(\alpha)=\frac{1}{2\pi\sigma_x\sigma_y}\,\Exp{-\frac{x^2}{2\sigma_x^2}-\frac{y^2}{2\sigma_y^2}},
\end{equation}
with $\alpha=x+iy$. For our purposes, it is more appropriate to define the Wigner function in a rotated plane with the rotation angle corresponding to the phase $\varphi$ of the local oscillator in a homodyne measurement. In the rotated frame, the Wigner function is 
\begin{equation}
W_\varphi(\alpha_\varphi)=W(\Exp{i(\varphi-\theta/2)}\alpha_\varphi-\moy{a}\Exp{-i\theta/2})\end{equation}
with  $\alpha_\varphi=x_\varphi+iy_\varphi$.
The marginal along $x_\varphi$, $P(x_\varphi)$, is obtained after integrating the Wigner function on $y_\varphi$:
\begin{align}
P(x_\varphi)&=\frac{1}{\sqrt{2\pi\sigma_\varphi}}\,\Exp{-\frac{(x_\varphi-x_0)^2}{2\sigma_\varphi^2}},\\
\sigma_\varphi^2&=\cos^2(\varphi-\theta/2)\sigma_x^2+\sin^2(\varphi-\theta/2)\sigma_y^2,\\
x_0&=\mathrm{Re}(\moy{a}\Exp{-i\varphi}).
\end{align}
In practice, we chose $\varphi=\theta/2$ in the squeezing direction such that the variance of the marginal is minimal.

In the presence of a dispersively coupled qubit, the field evolves to two distinct states.  We denote the marginals associated to the two states as $P_{j=0,1}(x)=\Exp{-(x-x_j)^2/2\sigma_j^2}/\sqrt{2\pi}\sigma_j$, with $x_1>x_0$.
When $\sigma_0\neq\sigma_1$, the two marginals intersect at the points 
\begin{multline}
z_{0,1}=\frac{1}{\frac{1}{\sigma_0^2}-\frac{1}{\sigma_1^2}}\left[\frac{x_0}{\sigma_0^2}-\frac{x_1}{\sigma_1^2}\right.\\\left.\mp\sqrt{\left(\frac{x_0}{\sigma_0^2}-\frac{x_1}{\sigma_1^2}\right)^2-\left(\frac{1}{\sigma_0^2}-\frac{1}{\sigma_1^2}\right)\left(\frac{x_0}{\sigma_0^2}-\frac{x_1}{\sigma_1^2}-2\ln\frac{\sigma_1}{\sigma_0}\right)}\right].
\end{multline}
The error on the distinguishability is quantified through the overlap between the two marginals, $E=\int\min(P_0,P_1)/2$~\cite{sete:2013a}.
The integration yields
\begin{multline}
E=\frac{1}{2}-\frac{1}{4}\left[\mathrm{erf}\left(\frac{z_+-x_1}{\sqrt{2}\sigma_1}\right)-\mathrm{erf}\left(\frac{z_--x_1}{\sqrt{2}\sigma_1}\right)\right.\\\left.-\mathrm{erf}\left(\frac{z_+-x_0}{\sqrt{2}\sigma_0}\right)+\mathrm{erf}\left(\frac{z_--x_0}{\sqrt{2}\sigma_0}\right)\right].
\end{multline}
When the two variances are equal, $\sigma_0=\sigma_1\equiv\sigma$, they cross at $z=(x_0+x_1)/2$.
The resulting error reads
\begin{equation}
E=\frac{1}{2}-\frac{1}{2}\mathrm{erf}\left(\frac{x_1-x_0}{2\sqrt{2}\sigma}\right).
\end{equation}

\bibliography{kmod.bbl}

\end{document}